\documentclass[aps, prd,twocolumn, notitlepage, nofootinbib]{revtex4-1}
\usepackage{graphicx}
\usepackage{float}
\usepackage{amsmath}
\usepackage{color}
\usepackage{tensor}
\usepackage{epstopdf}
\usepackage{xcolor}
\usepackage{hyperref}
\hypersetup{
    colorlinks,
    linkcolor={blue!50!black},
    citecolor={red!50!black},
    urlcolor={blue!80!black}
}

\usepackage{multibib}

\newcommand{\be}{\begin{equation}}
\newcommand{\ee}{\end{equation}}
\newcommand{\ba}{\begin{eqnarray}}
\newcommand{\ea}{\end{eqnarray}}

\newcommand{\beq}{\begin{equation}}
\newcommand{\eeq}{\end{equation}}
\newcommand{\beqa}{\begin{eqnarray}}
\newcommand{\eeqa}{\end{eqnarray}}
\newcommand{\nn}{\nonumber}

\begin{document}

\title{Holographic heat engines: general considerations and rotating black holes}

\author{Robie A. Hennigar}
\email{rhennigar@uwaterloo.ca}
\affiliation{Department of Physics and Astronomy, University of Waterloo, Waterloo, Ontario, Canada, N2L 3G1}

\author{Fiona McCarthy}
\email{fmccarthy@perimeterinstitute.ca}
\affiliation{Perimeter Institute, 31 Caroline St. N., Waterloo,
Ontario, N2L 2Y5, Canada}

\author{Robert B. Mann}
\email{rbmann@uwaterloo.ca}
\affiliation{Department of Physics and Astronomy, University of Waterloo, Waterloo, Ontario, Canada, N2L 3G1}
\affiliation{Perimeter Institute, 31 Caroline St. N., Waterloo,
Ontario, N2L 2Y5, Canada}

\author{Alvaro Ballon}
\email{aballonbordo@perimeterinstitute.ca}
\affiliation{Perimeter Institute, 31 Caroline St. N., Waterloo,
Ontario, N2L 2Y5, Canada}

%


\begin{abstract}

We perform the first study of holographic heat engines where the working material is a rotating black hole, obtaining exact results for the efficiency of a rectangular engine cycle.  We also make general considerations in the context of benchmarking these engines on circular cycles.  We find an exact expression that is valid for black holes with vanishing specific heat at constant volume and derive an upper bound, below the Carnot efficiency and independent of spacetime dimension, which holds for any black hole of this kind.  We illustrate our results with applications to a variety of black holes, noting the effects of spacetime dimension, rotation, and higher curvature corrections on the efficiency of the cycle.
\end{abstract}

\maketitle

\section{Introduction}

Thermodynamic aspects of black holes provide deep insights into the nature of quantum gravity.  The Bekenstein-Hawking entropy motivated the holographic principle~\cite{Susskind:1994vu}, while the thermodynamic behaviour of black holes in anti de Sitter (AdS) space sheds light on strongly coupled gauge theories through various proposed dualities. 

Recently, there has been growing interest in the subject of \textit{black hole chemistry}.  Here the cosmological constant is elevated to a thermodynamic parameter, appearing as a pressure in the first law of thermodynamics~\cite{Teitelboim:1985dp, Caldarelli:1999xj, Cvetic:2010jb, Dolan:2010ha}. This interpretation arises naturally via many mechanisms, e.g. from the coupling of gravity to gauge fields~\cite{Creighton:1995au}, and is supported by geometric arguments~\cite{Kastor:2009wy}.  From this identification, a number of interesting results have followed.  The thermodynamic volume, which is defined as the conjugate to the pressure, has been conjectured to satisfy a reverse isoperimetric inequality~\cite{Cvetic:2010jb, Hennigar:2014cfa} (which bounds the black hole entropy corresponding to a particular thermodynamic volume) that has recently been related to an upper bound on the holographic butterfly velocity~\cite{Feng:2017wvc}.  Interesting phase structure has emerged, including a direct physical analogy between the charged AdS black hole and the van der Waals fluid~\cite{Kubiznak:2012wp}, re-entrant phase transitions~\cite{Altamirano:2013ane, Frassino:2014pha}, isolated critical points~\cite{Dolan:2014vba}, and recently the analog of a superfluid phase transition~\cite{Hennigar:2016xwd}.  Numerous other interesting results have also been obtained and we refer the reader to~\cite{Kubiznak:2016qmn} for a recent review.
 
Within the context of black hole chemistry, it becomes natural to consider \textit{holographic heat engines}: cycles in the pressure-volume space that extract work from the AdS black holes used as the working material~\cite{Johnson:2014yja}.  The engines are named \textit{holographic} because, for a negative cosmological constant, the engine cycle corresponds to a process defined on the space of dual field theories in one dimension lower~\cite{Johnson:2014yja}.  A number of subsequent studies have deepened the holographic understanding of these processes in which the cosmological constant is adjusted~\cite{Caceres:2015vsa, Couch:2016exn, Caceres:2016xjz, Dolan:2016jjc, Karch:2015rpa}. 

Many interesting results have been obtained for black hole heat engines, among them being exact results for certain classes of static black holes~\cite{Johnson:2016pfa}, the effects of higher curvature corrections~\cite{Johnson:2015ekr} (which correspond to $1/N_c$ corrections in the dual CFT), and using engine cycles as a means to compare different black holes through a benchmarking prescription~\cite{Chakraborty:2016ssb}.  Similar results have been found and elaborated upon by other authors~\cite{Zhang:2016wek, Bhamidipati:2016gel, Wei:2016hkm, Setare:2015yra, Sadeghi:2015ksa}. More recently, Johnson has shown~\cite{Johnson:2017hxu} that an engine cycle taking advantage of a critical point can have efficiency approaching the Carnot efficiency despite not being a Carnot cycle, providing perhaps the first exact solution for such a result.  

To date, all considerations of holographic heat engines have relied on the black holes having vanishing specific heat at constant volume, $C_V = 0$.  Under this restriction, the thermodynamic equations have been found to be far more manageable.  However, as a consequence, the effects of rotation have been ignored, since the Kerr-AdS family of solutions has $C_V \neq 0$.  One aim of this work is to remedy this: we consider holographic heat engines without the restriction to $C_V = 0$ and show that exact results can be obtained for rectangular cycles expressed simply in terms of the mass and internal energy of the black hole.   

The second aim of our paper is to contextualize rotating black holes in the benchmarking scheme recently presented by Johnson and Chakraborty~\cite{Chakraborty:2016ssb}.  Here, we consider an elliptical cycle and show that the efficiency of this cycle can be evaluated exactly in the case $C_V = 0$.  In this case, we also show that the efficiency of any black hole with $C_V  = 0$ is bounded from above by $\eta_\circ = 2\pi/(\pi + 4) < \eta_C$, where $\eta_C$ is the Carnot efficiency.  This upper bound is attained by extremal black holes in the limit of a small cycle.  

We compare the efficiency of a variety of black holes, arriving at a number of interesting conclusions.  We find that the presence of non-linear electrodynamics tends to increase the engine efficiency.  In the limit of small or vanishing work terms (e.g. charge or angular momentum), the efficiency is found to depend sensitively on the topology of the event horizon, with hyperbolic black holes the most efficient, followed by planar and then spherical.  The super-entropic black hole, which has an event horizon which is topologically a sphere with two punctures~\cite{Hennigar:2014cfa, Klemm:2014rda}, is found to be the most efficient in this limit.  This result is intriguing since the super-entropic black hole is also the only known counter-example to the reverse isoperimetric inequality in Einstein gravity~\cite{Hennigar:2014cfa}, suggesting a possible link between the inequality and the engine efficiency.  The addition of rotation tends to lead to a decrease in efficiency for large angular momentum, but for certain intermediate values of the angular momentum the efficiency can attain values larger than that of a compareable non-rotating black hole.  We also consider higher curvature corrections of the Lovelock~\cite{Lovelock:1971yv} type and find that the efficiency of planar black holes is unaffected by the presence of these terms.  Corrections  due to even powers of the curvature serve to decrease the efficiency of engines that are spherical or hyperbolic in topology.  Corrections due to odd powers of the curvature serve to decrease the efficiency of spherical black holes, but lead to an increase in the efficiency for hyperbolic black holes.

\section{Exact results for rectangular cycles}

We begin by making some general considerations for holographic heat engines akin to those made previously~\cite{Johnson:2016pfa}, but without the restriction $C_V=0$.  We are particularly interested in computing the efficiency 
\be\label{effic}
\eta = \frac{W}{Q_H} = 1 - \frac{Q_C}{Q_H}
\ee
of a cycle in which there is  a net  input heat  flow $Q_H$, a net output heat flow $Q_C$, and a net output work $W$.
 Our starting point will be a rectangular cycle.  As we will see, in many ways a rectangular cycle is the most natural cycle to consider for all AdS black holes as it will generalize to an algorithm which allows for more complicated cycles to be computed numerically---or even exactly, in some cases.  

Let this rectangle be described by the $(V, P)$ coordinates  $V_L, V_R$ and $P_T, P_B$ respectively: the subscripts $T$ and $B$ refer to ``top" and ``bottom", while the subscripts $L, R$ refer to ``left" and ``right".  The four combinations $(V_i, P_j)$ then give the coordinates of the corners of the rectangle.

First, consider the heat flow along the isobars.  In this case, as noted in~\cite{Johnson:2016pfa}, working with the enthalpy is convenient.  We perform the following simple manipulations:
\begin{align} 
\delta Q &= TdS = dM - VdP
\nn\\
\Rightarrow Q_{isobar} &= \int_a ^b d M = M_b - M_a 
\end{align}
where we have made use of the first law of black hole mechanics and the fact that we are considering an isobar. Note that, in general, the first law will contain additional work terms, e.g. angular momentum or electric charge.  Our results apply to the cases where these quantities are held constant through the engine cycle.  The calculation works analagously for an isochore; we find
\begin{align} 
\delta Q &= TdS = dU + PdV
\nn\\
\Rightarrow Q_{isochore} &= \int_a ^b d U = U_b - U_a 
\end{align}
where 
\be 
U = M - PV  
\ee
is the relationship between the internal energy $U$ and the enthalpy (mass).  We
obtain
\be 
Q_{isochore} = M_b - M_a - V (P_b - P_a) \, .
\ee
Thus, simple thermodynamic considerations have led to exact expressions for the heat flow on isobars and isochores in terms of quantities which are known for the black hole.

  In the case of a rectangular cycle, we can use these results to obtain an exact formula which applies to all AdS black holes.  In terms of the corners of the rectangle, the heats read:
\begin{align}
Q_C &=  M(V_R, P_T) - M(V_L, P_B) - \Delta P V_R  
\nn\\
Q_H &=  M(V_R, P_T) - M(V_L, P_B)  - \Delta P V_L  
\end{align}
where we have defined $\Delta P = P_T - P_B$. One then obtains a remarkably simple expression for the efficiency of any rectangular cycle which can be manipulated to read
\be \label{rect-exact}
\eta = \frac{\Delta V \Delta P}{\Delta M_T  + \Delta U_L} 
\ee
where $\Delta V = V_R-V_L$ and
\begin{align} 
\Delta M_T &= M(V_R, P_T)  - M(V_L, P_T)  
\nn\\
\Delta U_L &= U(V_L, P_T) - U(V_L, P_B) \, .
\end{align}

We emphasize that the above expression for the efficiency of a rectangular cycle applies to any AdS black hole, as the only assumptions going into the result are that the first law of thermodynamics holds and, of course, that the cycle is carried out quasi-statically so that the computation of the heat is valid. There is  the further assumption that the heat flow $Q_H$ occurs along the top and $Q_C$ along the bottom, which holds for most black holes.  In cases where it does not hold,\footnote{The only example of this we have found is the super-entropic black hole, which has many unusual properties.} the exact result can still be obtained via the same prescription, with the $\Delta M_T \to \Delta M_B$ and $\Delta U_L \to \Delta U_R$.  This is the first exact expression for the efficiency valid for a rotating black hole presented in the literature to date. 

\begin{figure}[htp]
\centering
\includegraphics[width=0.45\textwidth]{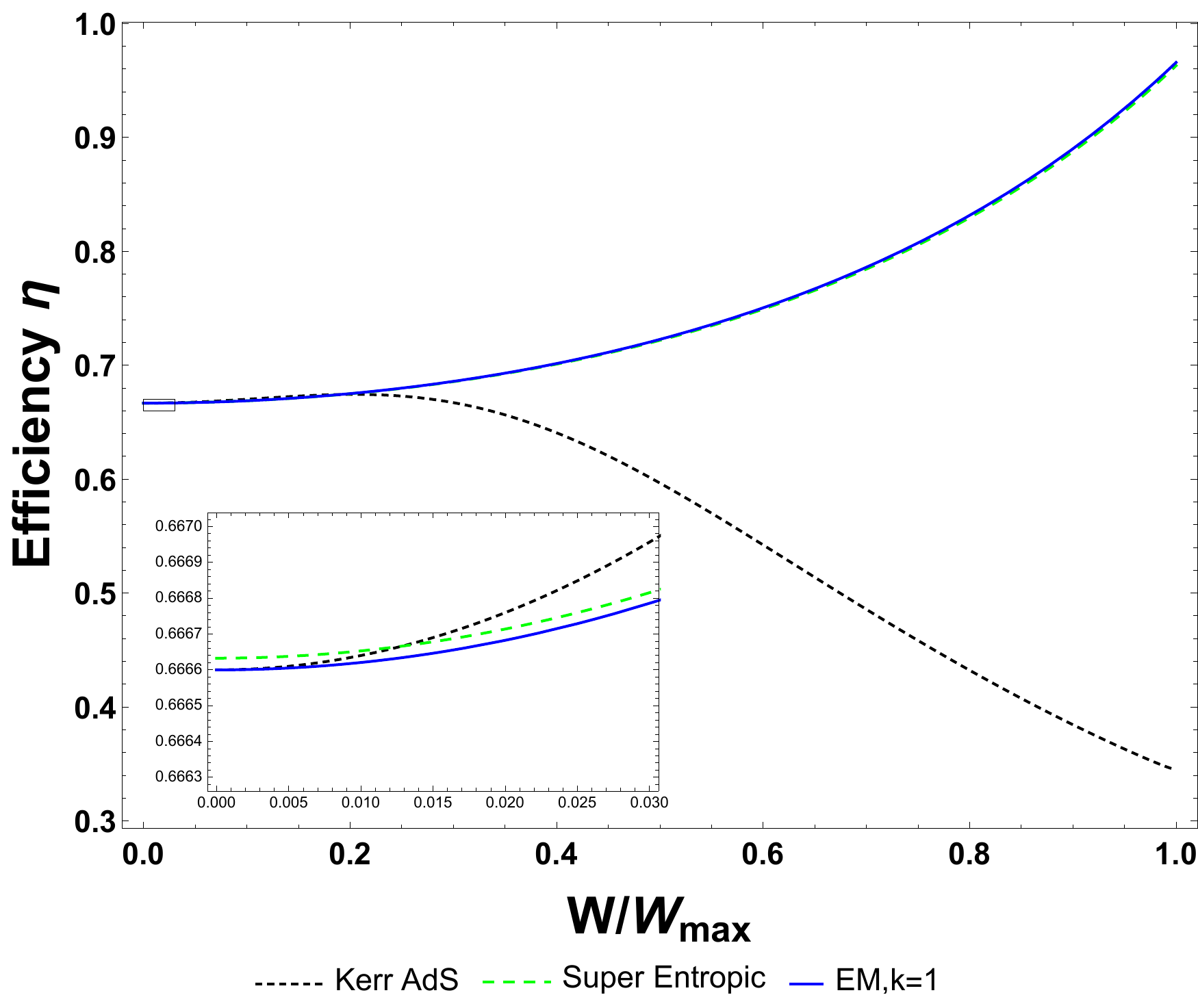}
\caption{{\bf Rectangular cycle} (color online).   A comparison of the Einstein-Maxwell black hole (solid, blue curve) with the Kerr-AdS (dotted, black curve) and charged super-entropic black hole (dashed, green curve) in four dimensions.  These latter two solutions have $C_V \neq 0$.  The quantity $W$ represents the work term (charge or angular momentum) with extremality occurring for $W = W_{\rm max}$. The cycle is centered at $(V_0, P_0) = (200, 20)$ with side lengths equal to $10$.}
\label{fig:rectangular_cycle}
\end{figure}

We present results for the efficiency in fig.~\ref{fig:rectangular_cycle} for two examples of black holes with $C_V \neq 0$: the Kerr-AdS solution and the charged  super-entropic black hole in four dimensions.  We have also included the Einstein-Maxwell black hole with spherical horizon topology for comparison. Plotting efficiency as a function of $W/W_{\rm max}$ with $W \in \{Q, J \}$ corresponding to either charge (Einstein-Maxwell and super-entropic\footnote{Note that the angular momentum of the super-entropic black hole is not a free parameter but is instead determined by the mass of the black hole.}) or angular momentum (Kerr-AdS) we see 
that for small to intermediate values of the angular momentum (relative to the extremal value) the efficiency for the Kerr-AdS black hole increases to exceed that of both the super-entropic and Einstein-Maxwell black holes.
  As the angular momentum of the Kerr-AdS black hole approaches its extremal value, the efficiency falls off, reaching a minimum at extremality.  On the other hand, the super-entropic black hole, for small charges, is more efficient than either the Kerr-AdS or Einstein-Maxwell black hole.  As the charge approaches the extremal value, the efficiency of the super-entropic black hole dips below the   the efficiency of the Einstein-Maxwell black hole, though the two efficiencies remain remarkably close together.    
  
It is interesting that while for the Kerr-AdS black hole, having $C_V \neq 0$ leads to a minimum in the efficiency for large angular momentum, the super-entropic black hole seems impervious to this.  We suspect that this could be due to the topological structure of the super-entropic black hole, whose horizon is topologically a sphere with two punctures.  As we will see in the following section, the efficiency of the engine is sesitive to the horizon topology of the black hole used as the working material.  A more interesting possibility would be that this effect is due to some as yet unexplored consequence of the reverse isoperimetric inequality applied to black hole heat engines, as the super-entropic black hole is currently the only known counter-example to this conjecture in Einstein gravity  for black holes asymptotic to AdS space~\cite{Cvetic:2010jb, Hennigar:2014cfa}.

In the case where $C_V = 0$, the exact result \eqref{rect-exact} for the rectangular cycle generalizes to an algorithm that can be used to determine the efficiency of an arbitrary cycle~\cite{Johnson:2016pfa}:  the arbitrary cycle is tiled with rectangular cycles, and the heat flows can be calculated noting that the only contributions will come from those segments of the rectangles intersecting the curve describing the engine cycle.  However, this is not true in the case where $C_V \neq 0$. Since the heat is not a state function  the infinitesimal heat flow $dQ = TdS$ along an arbitrary segment of the cycle cannot be split into independent heat flows along an isobar and isochore which join the ends of the segment.  Thus, the exact result for the rectangular cycle only generalizes to an algorithm when the cycle of interest is actually comprised of rectangles. However, the exact result is useful for gaining insight into the features of these black holes, and for more general cycles a numerical integration along the prescribed cycle can be done accurately and precisely.    

\section{Applications to benchmarking}

One application of this is to the benchmarking of black hole heat engines recently proposed by Chakraborty and Johnson~\cite{Chakraborty:2016ssb}.  The idea is to fix an engine cycle and compare the efficiency of different black holes used as the working substance for this engine. Since the efficiency is a dimensionless quantity, it can be used for meaningful comparisons here.  However, certain cycles may yield advantages for one black hole relative to another if the equation of state happens to be well adapted to the chosen fixed cycle.  Therefore, the cycle must be chosen judiciously. Chakraborty and Johnson~\cite{Chakraborty:2016ssb} have suggested using a circular/elliptical\footnote{Since both $P$ and $V$ are dimensionful quantities, the precise shape of the cycle will depend on the units these quantities are measured in.  So ``elliptical" is more technically correct than ``circular", though we will use both terms to refer to this cycle.} cycle for this purpose, since from point to point on the circle all thermodynamic quantities will be changing; this cycle should be equally difficult for all black holes.

It is also possible to compute the efficiency exactly and in full generality for the elliptical cycle.  However, the result is only of practical value in the case $C_V = 0$, as we shall elaborate upon below. Consider an ellipse in the $(V,P)$ plane given by the following parametric equation,
\begin{align}
P(\theta) &= P_0(1+  p \sin \theta )\, ,
\nn\\
V(\theta) &= V_0(1+  v \cos \theta )\, .
\end{align}
Here $p$ and $v$ are dimensionless and correspond to the size of the axes of the ellipse.  The key insight here comes from the exact expression for rectangular cycles. In the case where $C_V = 0$, we can consider tiling the circle with rectangles, from which it becomes clear that there will be a contribution to $Q_H$ from the top half of the circle and a contribution to $Q_C$ from the bottom half of the circle due to the limiting behaviour of heat flow along isobars. 

The heat flows can be  calculated using the parameterization introduced above and utilizing the first law.  For example, the contribution to $Q_C$ from the bottom of the circle is
\begin{align} 
Q_C  &= \Delta M - \int_{\theta = 0}^{\theta = \pi } V_0(1 + v \cos \theta)P_0p \cos \theta d\theta 
\nn\\
&= \Delta M - P_0 V_0 \frac{\pi p v}{2} \, ,
\end{align}
where 
\be\label{eqn:deltaMcircle} 
\Delta M = M\left(V_0 (1 + v), P_0\right) - M(V_0 (1 - v), P_0) \, .
\ee
Performing analogous calculations for the remaining contribution to the heat, the efficiency of the circular cycle for $C_V = 0$ is obtained:
\be\label{eqn:exact_circle}
\eta = \frac{2 \pi }{\pi + \frac{2}{pv} \frac{\Delta M }{P_0V_0}} 
\ee
where $\Delta M$ is the same as in Eq.~\eqref{eqn:deltaMcircle}.

The problem with the $C_V \neq 0 $ case comes from the fact that one does not know what the limits of integration are in determining the heat.  We were able to obtain these here by considering the limiting behaviour of the rectangular cycles used to tile the circle.  Since this algorithm is valid only when $C_V = 0$, we cannot use this to gain insight in the $C_V \neq 0$ case.  In practice, it is easier to integrate $TdS$ numerically than it is to determine the bounds of integration and then apply the exact result.

Let us now consider some interesting limiting behaviour of the exact $C_V = 0$ circular cycle.  First we note that in the case of the ideal gas black hole (for which $M = PV$), this expression reduces to 
\be 
\eta = \frac{2 \pi }{\pi + 4  / p} \, ,
\ee
which is identical to the expression given in~\cite{Chakraborty:2016ssb}.  This expression implies an upper-bound for the efficiency of an ideal gas black hole: for any elliptical cycle, the largest permissible value of $p$ is $p=1$, corresponding to $\eta \approx 0.8798$.

The exact expression we have found for the circular cycle with $C_V = 0$
makes it is easy to extract a lower bound on the efficiency,
\be 
\eta \ge \eta_{\rm min } = \frac{2 \pi }{\pi + 2 M_R/P_0 p V_0 v} 
\ee 
where $M_R = M(V_0 (1 + v), P_0)$.  Equality is obtained in the limit that $v=1$, and $\eta_{\rm min} \to 0$ in the limit $M_R/P_0 p V_0 v \to \infty$. Thus, for general cycles using black holes of positive mass, this provides a lower bound on the efficiency, which is greater than zero.

The expression~\eqref{eqn:exact_circle}  makes clear what is necessary to maximize efficiency on a circular cycle: for a given, fixed cycle, the black hole whose $\Delta M$ is the smallest will perform most efficiently.
For cycles having $v \ll 1$ we can write
\be 
\Delta M \approx 2 V_0 v \frac{\partial M}{\partial V} \bigg|_{V=V_0} + \frac{(V_0v)^3}{3} \frac{\partial^3  M}{\partial V^3} \bigg|_{V=V_0} + \cdots \, ,
\ee
with the derivatives taken at constant pressure. We can therefore conclude the following for black holes with $C_V = 0$: \textit{any quantity  contributing negatively to $\partial M / \partial V$ will lead to a larger efficiency}.  An example of this would be electric charge.

In fact, this expression suggests an upper bound for the efficiency of a circular cycle for a certain class of asymptotically AdS black holes with $C_V = 0$.  Working in the narrow cycle limit, i.e. $v \ll 1$, $\Delta M$ can be accurately approximated with the first term of the derivative expression. We can use the chain rule to gain some insight into this term:
\be\label{eqn:dmdv_expr} 
\frac{\partial M}{\partial V}  = \frac{\frac{\partial M}{\partial S}}{\frac{\partial V}{\partial S}}  = \frac{T}{\partial V/\partial S} \, .
\ee
The temperature will always satisfy $T \ge 0$ with equality only for extremal black holes, so we must only study $(\partial V / \partial S)_P$.  We can use a Maxwell relation followed by the cyclic identity to show that
\be 
\left(\frac{\partial V}{\partial S} \right)_P = \left(\frac{\partial T}{\partial P} \right)_S\, ,
\ee 
and hence
\be 
\frac{\partial M}{\partial V}  = \left[ \left(\frac{\partial} {\partial P} \right)_S \log T \right]^{-1} \, .
\ee
Now let us assume that the temperature at the bottom of the cycle is known and write that $T_- = T(V_0, P_0(1-p))$.  Then we can, for small cycles, expand the temperature in a Taylor series,
\be 
T(V_0, P_0) = T_- + P_0 p \frac{\partial T}{\partial P} + \frac{(P_0p)^2}{2}  \frac{\partial^2  T}{\partial P^2} + \cdots\, ,
\ee 
giving
\be 
\left[ \left(\frac{\partial} {\partial P} \right)_S \log T \right]^{-1}  =  \frac{T_- + P_0 p \frac{\partial T}{\partial P} + \frac{(P_0 p)^2}{2}  \frac{\partial^2  T}{\partial P^2} + \cdots}{\frac{\partial T}{\partial P} + P_0 p  \frac{\partial^2  T}{\partial P^2} + \cdots } \, .
\ee 
Let us now consider this expansion as $T_- \to 0$.  We also drop the subscript $S$ on the derivatives, noting that the derivatives of temperature are taken at constant entropy (or equivalently at constant volume since $C_V = 0$). We first assume that 
\be 
p \ll \frac{1}{P_0}\frac{\partial T/\partial P}{\partial^2  T/\partial P^2}\, .
\ee 

This allows us to conclude that the first two terms in the series expansion of the temperature provide a good approximation to the temperature at the center of the cycle, allowing us to write
\begin{align} 
\frac{\partial M}{\partial V} 
&\approx  \frac{\left[ T_- + P_0 p \frac{\partial T}{\partial P} + \frac{(P_0p)^2}{2}  \frac{\partial^2  T}{\partial P^2} \right]\left[ 1 - P_0 p  \frac{\partial^2  T/ \partial P^2}{\partial T/ \partial P} \right]}{\frac{\partial T}{\partial P}} \, .
\end{align} 
We now consider the low temperature limit for the cycle, i.e.
\be 
T_- \ll P_0 p \frac{\partial T}{\partial P} \, .
\ee
In this limit we can write
\be 
\frac{1}{P_0 p} \frac{\partial M}{\partial V} = 1 + \frac{T_-}{P_0 p (\partial T / \partial P)} + \cdots   \, ,
\ee
where $\cdots$ represent very sub-leading terms.  Therefore, the efficiency is given by
\be 
\eta = \eta_\circ \left[1 -  \frac{T_-}{P_0 p (\partial T / \partial P)} + \cdots  \right]\,,
\ee 
and we have $\eta \leq \eta_\circ$, where
\be 
\eta_\circ = \frac{2 \pi}{\pi + 4}
\ee 
and equality is attained in the extremal limit.  Note that this upper bound is \textit{theory independent}: extremal black holes should always approach this efficiency in the small cycle limit regardless of spacetime dimension or the theory from which the extremal black holes are derived.  The subleading terms, characterizing the approach to $\eta_\circ$,
\be 
\frac{T_-}{P_0 p (\partial T / \partial P)} + \cdots \,,
\ee
will depend on the theory under consideration.

In our argument, we have implicitly assumed that $\partial M/ \partial V > 0$, but is this a good assumption? To study this we can consider the denominator of Eq.~\eqref{eqn:dmdv_expr}:
\be 
\left(\frac{\partial V}{\partial S} \right)_P = \left(\frac{\partial T}{\partial P} \right)_S = \frac{\alpha_P V T}{C_P} \, .
\ee
In this expression, $T$ and $V$ should be positive quantities, while $C_P$ must be positive for equilibrium stability.  We cannot \textit{a priori} confirm the positivity of the coefficient of thermal expansion, $\alpha_P = \tfrac{1}{V} \left(\tfrac{\partial V}{\partial T} \right)_P$, but using the relationship between heat capacities,
\be 
C_P - C_V = \frac{VT \alpha_P^2}{\beta_T}\, ,
\ee
where $\beta_T = - (\partial V / \partial P)_T/V$ is the isothermal compressibility, we can draw useful conclusions.  When $C_V=0$, provided $V$ and $T$ are positive, as we are assuming, each term is positive from thermodynamic stability.  This indicates that the
coefficient of thermal expansion should be either strictly positive or strictly negative.  Thus, from these arguments, since
\be 
\left(\frac{\partial M}{\partial V} \right)_P  = \frac{C_P}{\alpha_P V} 
\ee
 we can conclude that this term will be either strictly positive or strictly negative.  Our expression for the efficiency, in the small cycle limit, only makes sense in the case that this term is positive, since otherwise the efficiency would be greater than one.  Hence, in situations where our expression for the efficiency of the circular cycle is valid, we can assume without loss of generality that $\partial M/\partial V > 0$.

To sum up, for circular cycles for black holes with $C_V = 0$, we have found two bounds between which the efficiency must lie,
\be 
\frac{2\pi}{\pi + 2 M_R/ P_0 pV_0 v} \le \eta \le \frac{2 \pi}{\pi + 4}\, ,
\ee 
where equality on the left side is obtained in the limit $v = V_0$ (i.e. the cycle is as large as possible). The right-hand-side inequality is universal: independent of both theory and spacetime dimension, with equality obtained for extremal black holes in the small cycle limit.  Note that in the case of some hyperbolic black holes, the mass $M_R$ may be negative~\cite{Mann:1997jb}. In these cases, the lower bound will be given by zero.  

It is tempting to justify this result using intuition from the Carnot cycle and conclude that the greatest efficiency will be obtained when at one point along the cycle the black hole is extremal.  However, when $C_V \neq 0$ this is not the case.  Additional heat flows occurring along isochores will in general result in a decrease in the efficiency of the engine.  
\begin{figure}[htp]
\centering
\includegraphics[width=0.42\textwidth]{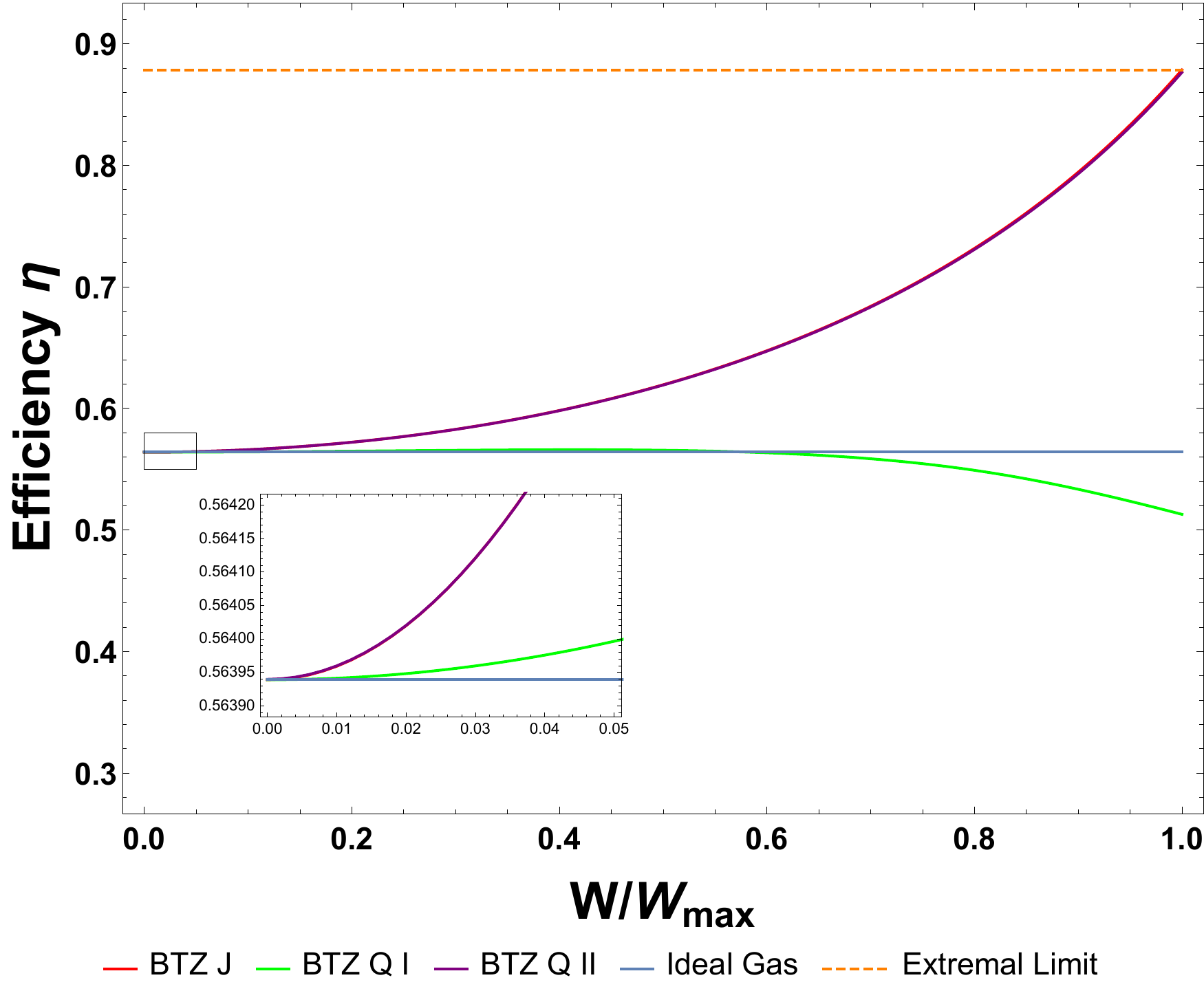}
\includegraphics[width=0.45\textwidth]{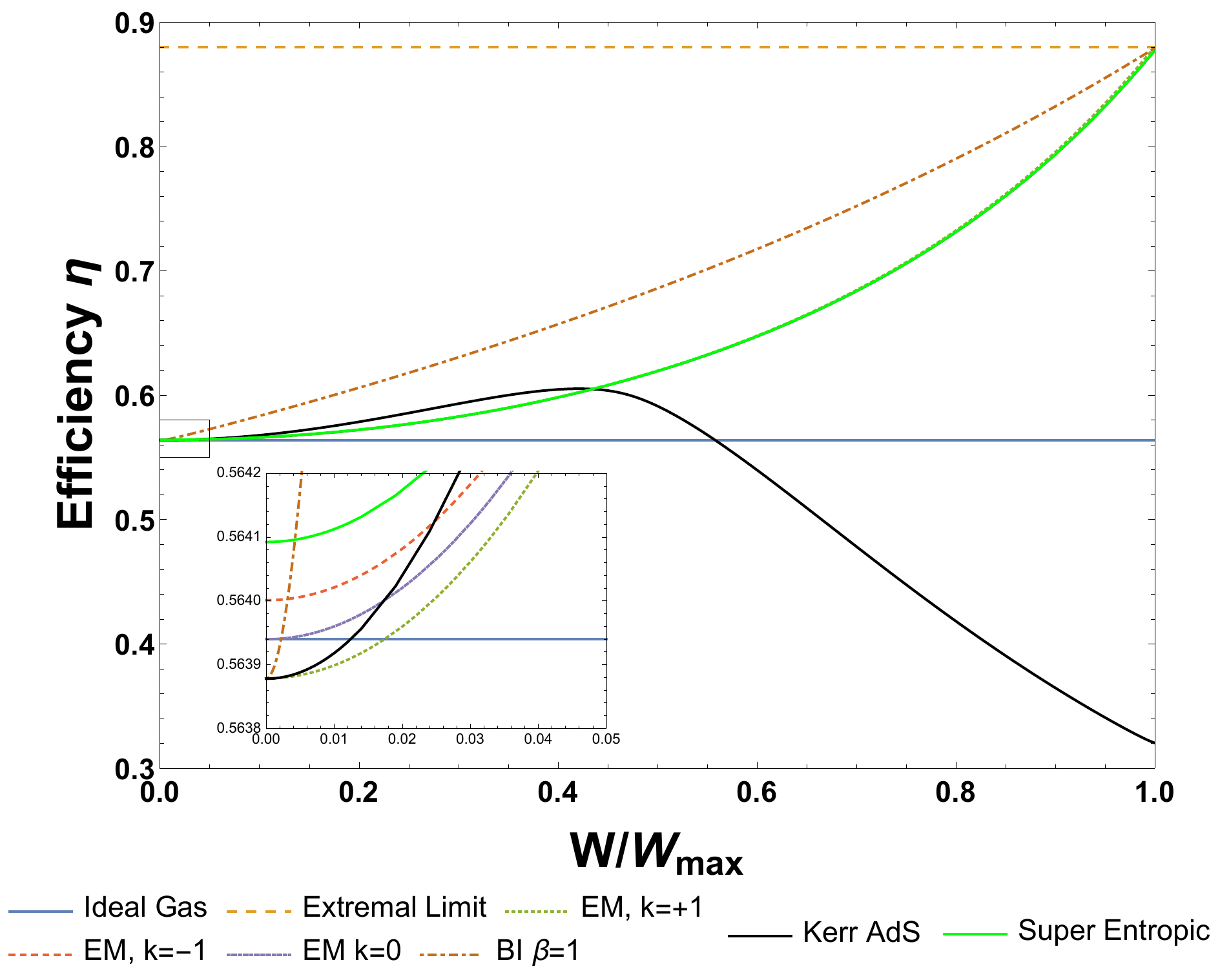}
\includegraphics[width=0.45\textwidth]{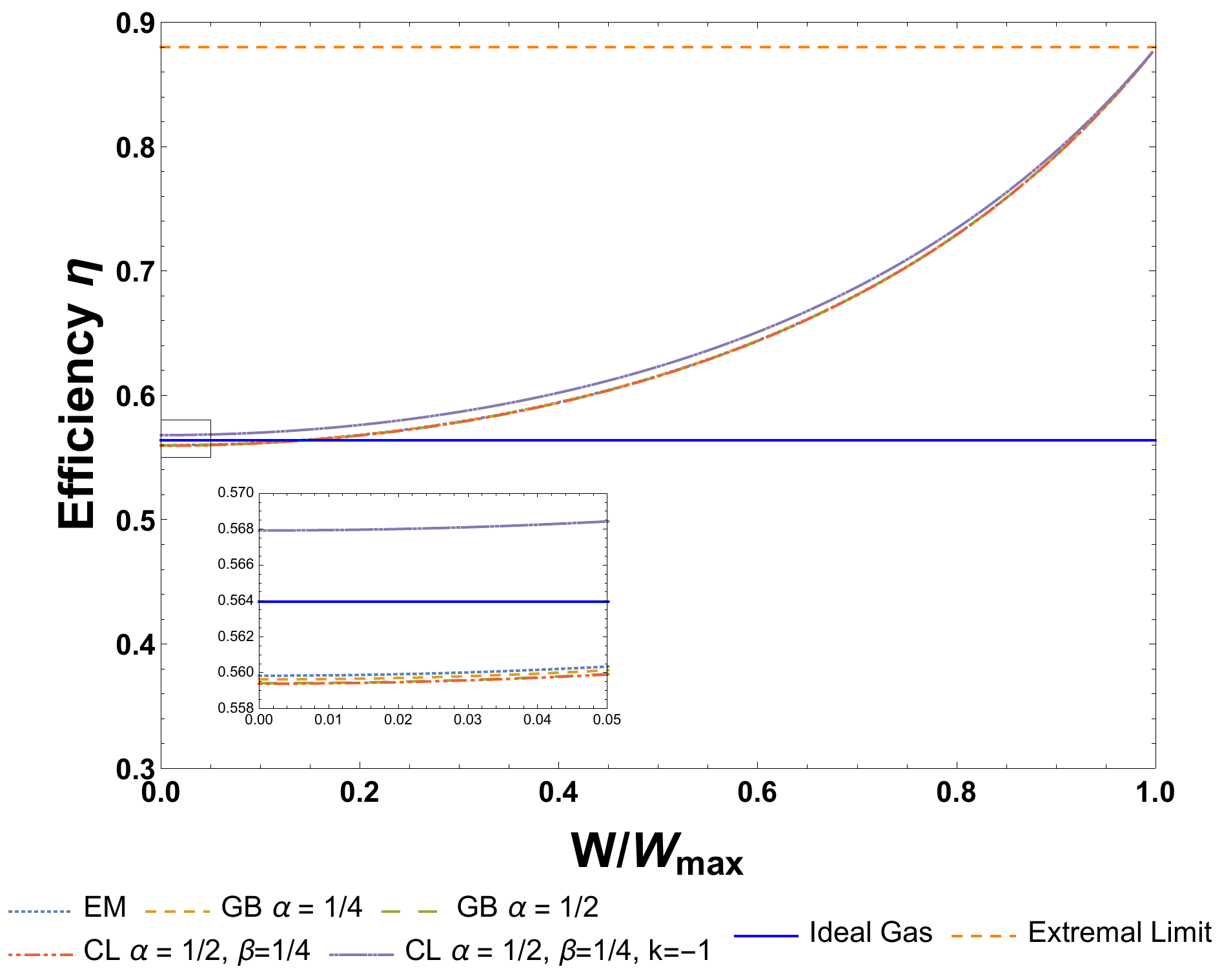}
\caption{{\bf Benchmarking curves} (color online). 
\textit{Top}: three dimensions.  \textit{Center}: four dimensions. \textit{Bottom}: eight dimensions.  In all cases, $W$ represents the work term (e.g. electric charge) and $W = W_{\rm max}$ gives an extremal black hole at one point on the cycle. In each case, the cycle was centered at $(V_0, P_0) = (200, 20)$.}
\label{fig:bench_marking}
\end{figure}

We conclude with a consideration of benchmarking of the black hole heat engines.  We present in fig.~\ref{fig:bench_marking} plots of benchmarking curves for various black hole heat engines in three, four and eight dimensions (top to bottom).  The metrics and relevant thermodynamic potentials for these black holes can be found in the appendix. In each plot we consider black holes with different work terms (charge or angular momentum) denoted by $W$ normalized by the value corresponding to an extremal black hole, denoted by $W_{\rm max}$. We include in each plot both the ideal gas case and the extremal limit we obtained above; each of these appear as horizontal lines.  In all cases where $C_V \neq 0$, we have performed a numerical integration to obtain the heat, crosschecking the results with the exact expression in the $C_V = 0$ case. We have confirmed that our numerical scheme is accurate at least to the number of decimal places presented in the plots.  

The plots reveal a number of interesting results. 
In three dimensions we show three curves for the BTZ black hole, comparing the rotating case to the charged case.  In the charged case, there is a subtlety in defining the thermodynamics depending on whether the AdS length is used or a new dimensionful quantity is introduced to make the horizon radius dimensionless inside a logarithm (see the discussion in~\cite{Frassino:2015oca} for more details).  The former case we label as ``BTZ Q I" while the latter is labeled as ``BTZ Q II".  When a new dimensionful quantity is introduced we have $C_V = 0$ and the benchmarking curve for the charged BTZ black hole is indistinguishable from that for the rotating BTZ black hole, with both approaching $\eta_\circ$ in the extremal limit and limiting to the ideal gas value when $W  \to 0$. When the AdS length is used, one has $C_V \neq 0$, and the resulting efficiency is always less than the corresponding results for the rotating BTZ black hole or the charged BTZ black hole using scheme II.  In the limit $W \to 0$, all three curves limit to the ideal case value. 

In the center plot of fig.~\ref{fig:bench_marking} we display various benchmarking curves for four dimensional black holes.  The efficiency of the black holes which have $C_V = 0$ all approach $\eta_\circ$ in the extremal limit. The presence of nonlinear Born-Infeld electrodynamics serves to increase the efficiency for the intermediate values of charge, $0 < Q < Q_{\rm max}$. We see that the Kerr-AdS solution, which does not have $C_V = 0$, reaches a peak in the efficiency near $J/J_{\rm max} \approx 0.5$ and then becomes rapidly less efficient as the extremal limit is approached.  We have not been able to determine any universal features pertaining to the location of this maximum; its location and height can depend on the details of the benchmarking cycle chosen.  The charged super-entropic black hole, which also has $C_V \neq 0$, displays quite interesting behaviour: the benchmarking curve follows closely those for the black holes with $C_V = 0$, and  in the extremal limit the efficiency approaches $\eta_\circ$.  

When the work term is small compared to its maximum value, ($W/W_{\rm max} \ll 1$) we can see some interesting structure in way the the black holes are ordered by efficiency, as shown in the inset of the center plot of fig.~\ref{fig:bench_marking}.  Consider first  topological black holes: the hyperbolic $k=-1$ case is most efficient, followed by the $k=0$ planar and then the $k=1$ spherical black holes.  These results are consistent with the statement made earlier: a quantity that contributes negatively to  $\partial M/\partial V$  will increase the efficiency. The topology enters into the mass in the form of $k$ multiplied by a positive quantity, explaining the observed behaviour.  It is interesting to note the super-entropic black hole is the most efficient in the $W \to 0$ limit, and we have confirmed that this is not a numerical error.  This is somewhat surprising and   is perhaps related to the horizon topology of these black holes being a sphere with two punctures. It also suggests a possible bound on the efficiency from the reverse isoperimetric inequality, at least in certain limiting cases.

The lower plot in fig.~\ref{fig:bench_marking} show various benchmarking curves for eight dimensional black holes.  We have chosen eight dimensions to highlight the effect of higher curvature corrections to the efficiency of the cycles. From the dual field theory perspective, the higher curvature terms serve as $1/N_c$ corrections. In~\cite{Johnson:2015ekr} Johnson observed that Gauss-Bonnet corrections lead to a decrease in the efficiency of an engine cycle for spherical black holes.  In general, the effect of higher curvature corrections depends sensitively on both the horizon topology and the sign of the higher curvature coupling constants.  Restricting attention to charged black holes in Lovelock theory~\cite{Lovelock:1971yv} (which have $C_V = 0$) and assuming the Lovelock couplings are positive, we can draw the following conclusions.  The Lovelock corrections  to the mass of order $K$ in the curvature, which become non-trivial in $d= 2K+1$, are of the form\footnote{See, e.g. Eq.~(2.19) in \cite{Frassino:2014pha}.} 
\be 
M_K \sim \alpha_K k^K r_+^{d - 2K - 1}\, .
\ee 
Here $k$ characterizes the geometry of the event horizon, with $k=+1,0,-1$ yielding spherical, planar and hyperbolic black holes, respectively.  The contributions from even $K$ will always serve to decrease the efficiency of the engine provided the topology is spherical or hyperbolic, since in these cases the contribution to $\partial M/\partial V$ is positive.  The contributions from odd $K$ will decrease the efficiency if the topology is spherical, but will increase the efficiency if the topology is hyperbolic.  The higher curvature corrections will have no effect on the efficiency for planar black holes.  Furthermore, we note that in the dimension $d = 2K + 1$, the $K^{th}$ order contributions will have no effect on the efficiency, since then there is no contribution to $\partial M/\partial V$.  We highlight some of these results for eight dimensional black holes in  Einstein-Maxwell, Gauss-Bonnet, and cubic Lovelock gravity in fig.~\ref{fig:bench_marking}.

\section{Conclusions}

We have presented a number of results for holographic heat engines where the working material is a rotating black hole, or---more generally---a black hole with non-vanishing $C_V$.  In particular, we have derived an exact result that applies to any black hole for a rectangular cycle.  This result generalizes to an algorithm  allowing one to determine exactly the efficiency of any cycle that can be decomposed exactly into rectangular pieces. We have found that, in general, rotation leads to a decrease in the efficiency of the engine.  However, for certain values of angular momentum (relative to the extremal angular momentum), the efficiency of the rotating black hole engine can approach---or even exceed---the efficiency of an analogous engine using a  non-rotating black hole.

Additionally, we have made some general considerations for benchmarking black hole heat engines.  We have found, in the limit of small cycles, an upper bound that heat engines exploiting black holes with $C_V = 0$ will satisfy:
\be
\eta < \eta_\circ = \frac{2 \pi}{\pi + 4} < \eta_C \, .
\ee
This upper bound, which resides below the Carnot bound and is independent of spacetime dimension, is saturated by extremal black holes. We have created various benchmarking curves for black holes in three, four and eight dimensions, observing them to obey this bound.  

In three dimensions, we studied the BTZ black hole under the two proposals for a thermodynamic description made in~\cite{Frassino:2015oca}.  We find that in the case where $C_V \neq 0$ (which occurs for the charged BTZ under certain thermodynamic identifications) the efficiency is decreased.   

In four dimensions a comparison of the topological black holes of Einstein-Maxwell theory, the Born-Infeld black hole, as well as the Kerr-AdS and super-entropic black holes, indicates that non-linear electrodynamics leads to an increase in   efficiency. Furthermore, we find that the horizon geometry affects the efficiency, with hyperbolic black holes being the most efficienct, followed by planar and then spherical black holes.  For the Kerr-AdS black hole, as the angular momentum approaches the extremal value, the efficiency reaches a minimum.  The super-entropic black hole displays interesting behaviour, being more efficient than even the hyperbolic black holes in the limit of small charge.  We suspect this either occurs due to the unique topology of the super-entropic black hole (which has horizon topologically a sphere with two punctures), or suggests a possible link between the reverse isoperimetric inequality and the efficiency of black hole heat engines (at least in certain limits, e.g. for small cycles).  This topic merits further investigation.

Exploring the effects of higher curvature corrections to the efficiency (which amounts to studying $1/N_c$ corrections in the dual field theory), we find the effect depends on the horizon topology.  Restricting to charged black holes of the Lovelock class, we found that planar black holes are uneffected by the addition of higher  curvature  terms.  Even-order Lovelock corrections decrease the efficiency of engines using spherical or hyperbolic black holes, while odd-order Lovelock corrections can actually increase the efficiency provided the black holes have hyperbolic horizons.  We also noted that $K^{th}$-order Lovelock terms in $d = 2K+1$ dimensions will not affect the efficiency. 

It would be worthwhile to extend these results to other cases; for example, higher dimensional black holes with multiple rotation parameters.  In higher dimensions, one could also make considerations similar to those of Johnson in~\cite{Johnson:2017hxu}: that is, one could determine, analytically, the efficiency of a heat engine near a critical point to determine if the rotation leads to any interesting deviations from the result when $C_V = 0$.  We leave these problems for future work.

\section*{Acknowledgments}
This work was supported in part by the Natural Sciences and Engineering Research Council of Canada.  We are grateful to the support
of the Perimeter Institute Winter School where this project was initiated.

\appendix

\section{Black hole metrics}

In this appendix we list, for convenience, the various black hole metrics we have used throughout this work and the thermodynamic parameters required to reproduce the calculations of this paper. In all cases, we normalize the cosmological constant in the standard way,
\be 
\Lambda = - \frac{(d-1)(d-2)}{2 l^2}\, ,
\ee
and the pressure is given by
\be 
P = - \frac{\Lambda}{8 \pi} \, .
\ee

\subsection{Three dimensions}

We use the conventions of~\cite{Frassino:2015oca} in the below metrics. 

\subsubsection{Rotating BTZ black hole}

The rotating BTZ black hole is given by the following metric:
\be 
ds^2 = - f dt^2 + \frac{dr^2}{f} + r^2 \left(d\varphi - \frac{J}{2 r^2} dt \right)^2\, ,
\ee
with
\be 
f = -2m + \frac{r^2}{l^2} + \frac{J^2}{4 r^2}\, ,
\ee
and has the following thermodynamic quantities:
\begin{align}
S &= \frac{\pi}{2} r_+ \, , \quad T = \frac{r_+}{2 \pi l^2} - \frac{J^2}{8 \pi r_+^3} \,, 
\nn\\
\Omega &= \frac{J}{16 r_+^2} \, , \quad M = \frac{m}{4} = \frac{r_+^2}{8 l^2} + \frac{J^2}{32 r_+^2} \, ,
\nn\\
P &= \frac{1}{8 \pi l^2} \, , \quad V = \pi r_+^2 \, .
\end{align}

\subsubsection{Charged BTZ black hole}

The metric for the charged BTZ black hole is given by
\be 
ds^2 = -f dt^2 + \frac{dr^2}{f} + r^2 d\varphi^2 \, ,
\ee
with
\be 
f = -2m + \frac{r^2}{l^2} - \frac{Q^2}{2} \log \left( \frac{r}{R} \right) \, .
\ee
Here $R$ is a constant with units of length.  The choice of this parameter determines whether BTZ scheme I or II is used.  If $R = l$ is  chosen, the this corresponds to BTZ I, and the thermodynamic quantities are
\begin{align} 
T &= \frac{r_+}{2 \pi l^2} - \frac{Q^2}{8 \pi r_+} \, , \quad S = \frac{\pi}{2} r_+\, , \Phi = - \frac{Q}{8} \log \left( \frac{r_+}{l} \right) \, ,
\nn\\
M &=  \frac{m}{4} = \frac{r_+^2}{8 l^2} - \frac{Q^2}{16} \log \left(\frac{r_+}{l} \right) \, , \quad V = \pi r_+^2 - \frac{\pi}{4}  Q^2 l^2 \, .
\nn\\
\end{align}
If instead $R$ is left as an independent, dimensionful parameter (motivated by the renormalization length introduced in the computation of the mass for this black hole), then this corresponds to BTZ II with the following thermodynamic parameters:
\begin{align} 
T &= \frac{r_+}{2 \pi l^2} - \frac{Q^2}{8 \pi r_+} \, , \quad S = \frac{\pi}{2} r_+\, ,
\nn\\
M &=  \frac{m}{4} = \frac{r_+^2}{8 l^2} - \frac{Q^2}{16} \log \left(\frac{r_+}{R} \right) \, , \quad V = \pi r_+^2\, .
\nn\\
K &= - \frac{Q^2}{16 R} \, , \quad  \Phi = - \frac{Q}{8} \log \left( \frac{r_+}{R} \right) \, ,
\end{align}
where $K$ is the thermodynamic conjugate to $R$.

\subsection{Four and higher dimensions}

\subsubsection{Born-Infeld black hole}

The metric for the Born-Infeld charged black hole takes the form~\cite{Johnson:2015fva}
\be 
ds^2 = -f dt^2 + \frac{dr^2}{f} + r^2 d \Omega_{d-2}^2\, ,
\ee
with 
\begin{align}
f &= 1 - \frac{m}{r^{d-3}} + \frac{r^2}{l^2}
\nn\\
 &+ \frac{4 \beta^2 r^2}{(d-1)(d-2)} \left[1 - \sqrt{1 + \frac{(d-2)(d-3)q^2}{2 \beta^2 r^{2d - 4}}} \right]
\nn\\
&+ \frac{2(d-2) q^2}{(d-1)r^{2d-4}} \times
\nn\\
&\times {}_2F_{1} \left[\frac{d-3}{2d-4}, \frac{1}{2}, \frac{3d - 7}{2d - 4}, - \frac{(d-2)(d-3)q^2}{2 \beta^2 r^{2d-4}} \right]\, ,
\end{align}
where ${}_2F_1$ is the hypergeometric function and $\beta$ is the Born-Infeld parameter.  When $\beta \to \infty$ this solution limits to the Einstein-Maxwell-AdS black hole.  The relevant thermodynamic quantities for the analysis in this paper are
\begin{align}
M &= \frac{(d-2) \omega_{d-2}}{16 \pi} m
\, , \quad V = \frac{\omega_{d-2}}{d-1} r_+^{d-1} \, ,
\nn\\
T &= \frac{f'(r_+)}{4 \pi} \, .
\end{align}

\subsubsection{Charged Lovelock black holes}
The metric for charged black holes in $K^{th}$ order Lovelock gravity take the form~\cite{Frassino:2014pha}
\be 
ds^2 = -f dt^2 + \frac{dr^2}{f} + r^2 d \Omega_{d-2}^2\, ,
\ee
with $f$ given by solving the polynomial equation
\begin{align}
\sum_{n=0}^{n_{max}} \alpha_n \left( \frac{k-f}{r^2} \right)^n =& \frac{16 \pi M}{(d-2) \omega_{k, d-2} r^{d-1}} 
\nn\\
&- \frac{8\pi Q^2}{(d-2)(d-3)} \frac{1}{r^{2d-4}}
\end{align}
where $n_{max}$ is the integer part of $(d-1)/2$.  For $n_{max} = 1$ or, equivalently, $\alpha_n = 0$ for $n > 1$, this gives the Einstein-Maxwell-AdS solution.  In the above, the following definitions are typically made:
\be 
\alpha_0 = \frac{1}{l^2} \, , \quad \alpha_1 = 1 \, .
\ee
The relevant thermodynamic quantities for the computations performed in this paper are
\begin{align}
M &= \frac{(d-2) \omega_{k, d-2}}{16 \pi} \sum_{n=0}^{n_{max}} \alpha_n k^n r_+^{d-1-2n} + \frac{\omega_{k, d-2}}{2(d-3)} \frac{Q^2}{r_+^{d-3}} \, , 
\nn\\
V &= \frac{\omega_{k, d-2} r_+^{d-1}}{d-1} \, , \quad
T = \frac{f'(r_+)}{4 \pi} \, ,
\end{align} 
and the remaining quantities can be found in~\cite{Frassino:2014pha}.

\subsubsection{Kerr-AdS solution}

The Kerr-AdS solution in four dimensions takes the form
\ba\label{KNADS}
ds^2&=&-\frac{\Delta_a}{\Sigma_a}\left[dt-\frac{a\sin^2\!\theta}{\Xi}d\phi\right]^2
+\frac{\Sigma_a}{\Delta_a} dr^2+\frac{\Sigma_a}{S}d\theta^2\nonumber\\
&&+\frac{S\sin^2\!\theta}{\Sigma_a}\left[a dt-\frac{r^2+a^2}{\Xi}d\phi\right]^2\,,
\ea
where
\ba
\Sigma_a&=&r^2+a^2\cos^2\!\theta\,,\quad \Xi=1-\frac{a^2}{l^2}\,,
\quad S=1-\frac{a^2}{l^2}\cos^2\!\theta\,,\nonumber\\
\Delta_a&=&(r^2+a^2)\Bigl(1+\frac{r^2}{l^2}\Bigr)-2mr \, .
\ea
The relevant thermodynamic parameters for this metric are~\cite{Caldarelli:1999xj, Hennigar:2014cfa} (see also~\cite{Dolan:2011jm})
\begin{align}
M &= \frac{m}{\Xi^2} \, , \quad J = \frac{ma}{\Xi^2} \, , \quad S = \frac{\pi(r_+^2 + a^2)}{\Xi}\, ,
\nn\\
V& =\frac{2\pi}{3}\frac{(r_+^2+a^2)(2r_+^2l^2+a^2l^2-r_+^2a^2)}{l^2\Xi^2r_+}
\nn\\
T &= \frac{r_+ \left(1 + \tfrac{a^2}{l^2} + 3 \tfrac{r_+^2}{l^2} - \tfrac{a^2}{r_+^2} \right)}{4 \pi (r_+^2 +a^2)}\, .
\end{align}

\subsubsection{Super-entropic black hole}

The charged super-entropic black hole in four-dimensions has the following metric~\cite{Hennigar:2014cfa}:
\ba\label{KNADS2}
ds^2&=&-\frac{\Delta}{\Sigma}\left[dt-l\sin^2\!\theta d\psi\right]^2
+\frac{\Sigma}{\Delta} dr^2+\frac{\Sigma}{\sin^2\!\theta}d\theta^2\nonumber\\
&&+\frac{\sin^4\!\theta}{\Sigma}\left[l dt-(r^2+l^2)d\psi\right]^2\,,\nonumber\\
{\cal A}&=&-\frac{qr}{\Sigma}\left(dt-l\sin^2\!\theta d\psi\right)\,,
\ea
where
\ba
\Sigma=r^2+l^2\cos^2\!\theta\,,\quad \Delta=\Bigl(l+\frac{r^2}{l}\Bigr)^2-2mr+q^2\,.\quad
\ea
with the following thermodynamic quantities:
\begin{align}
M&=\frac{\mu m}{2\pi}\,,\quad J=Ml\,,\quad 
\Omega = \frac{l}{r_+^2+l^2}\,,  \nonumber
\\
T&=\frac{1}{4\pi r_+}\left(3\frac{r_+^2}{l^2}-1-\frac{q^2}{l^2+r^2} \right), \nonumber \\ 
S&= \frac{\mu}{2}(l^2+r_+^2)=\frac{A}{4},\quad \Phi=\frac{qr_+}{r_+^2+l^2}, \;\;\; Q=\frac{\mu q}{2\pi}\, ,
\label{eq:thermo_properties}
\end{align}
where $\mu$ is the compactification parameter, i.e. the coordinate $\psi$ is indentified according to $\psi \sim \psi + \mu$.

\bibliography{mybib}
\end{document}